\documentclass[%
reprint,
superscriptaddress,
 amsmath,amssymb,
 aps,
]{revtex4-2}

\usepackage{graphicx}
\usepackage{hyperref}
\usepackage{dcolumn}
\usepackage{bm}
\usepackage{braket}
\usepackage{blindtext}
\usepackage{ragged2e} 
\usepackage{lipsum}

\usepackage{siunitx}
\DeclareSIUnit{\torr}{Torr}
\DeclareSIUnit{\mV}{mV}
\DeclareSIUnit{\mT}{mT}
\DeclareSIUnit{\nm}{nm}
\DeclareSIUnit{\rpm}{rpm}
\DeclareSIUnit{\sccm}{sccm}
\DeclareSIUnit{\angstrom}{\text{\AA}}
\DeclareSIUnit{\mK}{mK}
\DeclareSIUnit{\mum}{\micro\meter}

\usepackage{xcolor}
\definecolor{emily}{HTML}{5F00BA}
\definecolor{mario}{HTML}{5250FF}

\DeclareUnicodeCharacter{2212}{-}

\begin{document}

\preprint{APS/123-QED}

\title{Tunable inter-edge interactions in a bilayer graphene quantum Hall antidot}

\author{Mario Di Luca}
\author{Emily Hajigeorgiou}
\author{Zekang Zhou}
\author{Tengyan Feng}
\affiliation{Institute of Physics, École Polytechnique Fédérale de Lausanne (EPFL), CH-1015 Lausanne, Switzerland}

\author{Kenji Watanabe}
\affiliation{Research Center for Functional Materials, National Institute for Materials Science, 1-1 Namiki, Tsukuba 305-0044, Japan}

\author{Takashi Taniguchi}
\affiliation{International Center for Materials Nanoarchitectonics,
National Institute for Materials Science, 1-1 Namiki, Tsukuba 305-0044, Japan}

\author{Ferdinand Kuemmeth}
\affiliation{Institute of Experimental and Applied Physics and Halle-Berlin-Regensburg Cluster of Excellence CCE, University of Regensburg, Regensburg, Germany}

\author{Mitali Banerjee}
\email{mitali.banerjee@epfl.ch}
\affiliation{Institute of Physics, École Polytechnique Fédérale de Lausanne (EPFL), CH-1015 Lausanne, Switzerland}
\affiliation{Center for Quantum Science and Engineering (QSE Center), École Polytechnique Fédérale de Lausanne (EPFL), CH-1015 Lausanne, Switzerland}

\begin{abstract}
Quantum Hall antidots provide a promising geometry for probing quasiparticle transport and interference in the quantum Hall regime. Unlike conventional Fabry–Pérot interferometers, whose confined geometry may lead to Coulomb-dominated behavior, antidots offer access to both controlled quasiparticle localization and anyonic interference. In this letter, we investigate a gate-defined bilayer graphene antidot and demonstrate a tunable crossover between distinct transport regimes. By varying the antidot potential, we control the coupling between extended quantum Hall edge states and states localized around the antidot. For filling factor $\nu = 2$ and 4 this evolution is accompanied by a doubling of the conductance oscillation period, and an evolution of the magnetic field period from $\phi_0/\nu$ to $\phi_0$ which we identify with a crossover from a Coulomb-dominated regime to an Aharonov–Bohm regime in which transport is governed by interference around the antidot. The observed crossover reveals the importance of coupling between the antidot and extended edge states and establishes gate-defined antidots as a controllable platform for quantum Hall interferometry.

\end{abstract}

\maketitle

Electronic interferometers in the quantum Hall regime exploit chiral one-dimensional edge channels to probe the statistical exchange properties of emergent quasiparticles. The most common realization is the dot-like Fabry-P\'erot interferometer (FPI)~\cite{Halperin2011Apr}, used extensively in GaAs and graphene in the integer~\cite{McClure2009Nov, Ofek2010Mar, Deprez2021May, Ronen2021May, Fu2023Jan} and fractional quantum Hall regimes~\cite{Nakamura2019Jun, Nakamura2020Sep, Kim2024Nov, Werkmeister2025Apr, Samuelson2026Mar, Kim2026Jan}. However, the dot-like FPI can be affected by bulk-to-edge coupling and bulk impurities that can influence these measurements~\cite{Samuelson2025Jun}: when interfering edge states couple strongly to compressible bulk states, this coupling can dominate the response and mask the expected Aharonov-Bohm (AB) oscillations~\cite{Halperin2011Apr}. This has motivated alternative geometries free of bulk-to-edge coupling, such as two-path~\cite{Ji2003Mar, Neder2007Jul, Kundu2023Apr} and chiral~\cite{Deviatov2012} Mach-Zehnder interferometers.

Quantum Hall antidots (ADs), where interference occurs around a potential hill, offer another route free of bulk-to-edge coupling~\cite{LevySchreier2016Aug, Sim2008Feb}. Recent experiments have used ADs to localize quasiparticles in the fractional regime~\cite{DiLuca2026Aug_Nat, DiLuca2026Aug} and to generate controllable phase jumps in bilayer graphene FPIs~\cite{Henzinger2026Mar, Kim2026Mar} and chiral Mach-Zehnder interferometers~\cite{Ghosh2025Jun, Ghosh2025Jul}, operating the AD in the Coulomb-dominated regime as a tool to measure quasiparticle charge and electron temperature. The AD can, however, also be tuned into an AB regime with the advantage of being free from bulk-to-edge interactions.

In this letter, we use a dual-gate-defined bilayer graphene antidot, fully electrostatically controlled, to study the transition between these interaction regimes. By varying the antidot potential we tune inter-edge coupling, moving the system from a weakly coupled to a strongly coupled regime. In the weakly coupled regime we observe Coulomb-dominated oscillations at integer filling factor $\nu$, with a magnetic field period corresponding to a flux of $\phi_0/\nu$ and a gate voltage period set by the electron tunneling charge. In the strongly coupled regime, at even integer filling factors, the magnetic field periodicity changes from $\phi_0/\nu$ to $\phi_0$ and the gate voltage periodicity doubles, which we argue signals a transition from the Coulomb-dominated to the AB regime.\\

\begin{figure}[tb!]
 \includegraphics[width = 0.47\textwidth]{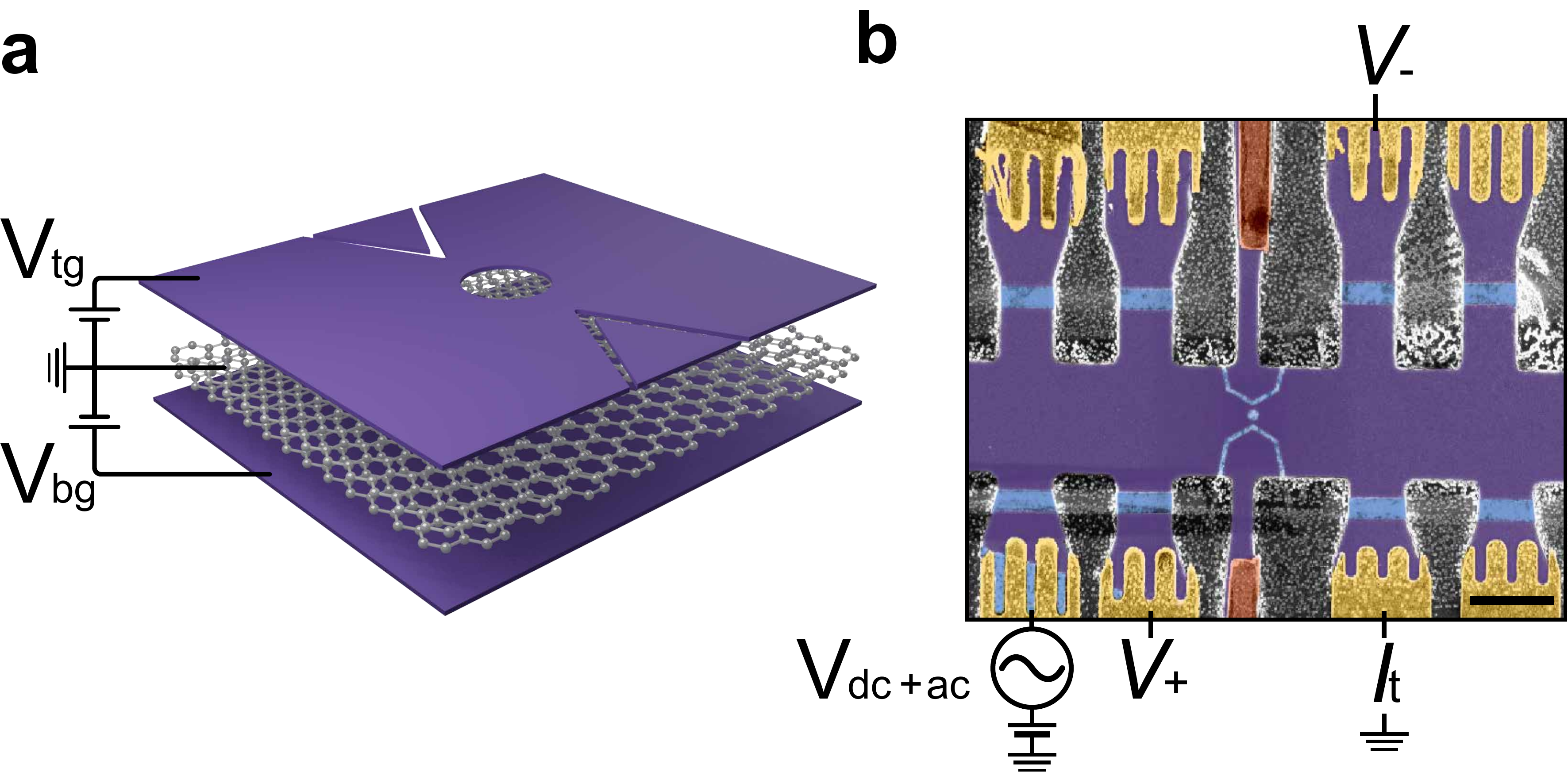}
 \begin{center}
 \caption{\textbf{Gate-defined antidot in bilayer graphene.} \textbf{a}, Schematic of the gate-defined antidot in bilayer graphene. The bilayer graphene is shown in gray, the top and bottom graphite gates are shown in purple, while the top and bottom hBN are not shown for clarity. Voltages $V_\mathrm{tg}$ and $V_\mathrm{bg}$ are applied to the TG and BG, respectively. \textbf{b}, False-color SEM image of an antidot device. The ohmic contacts are shown in gold, the palladium contacts to the graphite top gate in orange, the graphite in purple, and in blue the top hBN is visible in places where the top graphite has been etched. The AD area is defined by the etched circle in the top graphite. A voltage of V$_\mathrm{ac} = 10$ $\mu$V was applied, with measurement of the transmitted ($I_\mathrm{t}$) current. Voltage was measured immediately before the antidot ($V_\mathrm{+}$), and after the drain ($V_\mathrm{-}$). The scale bar is 2 $\mu$m.}
 \label{fig:device}
 \end{center}
\end{figure}

The AD is electrostatically defined by an etched hole in the top graphite, as shown in the schematic in Fig.~\ref{fig:device}a, with its density tuned near the highly resistive $\nu = 0$ state via a negative voltage applied to the bottom graphite gate (BG). The BLG layer is encapsulated between two layers of hexagonal boron nitride (hBN). A scanning electron micrograph (SEM) of the device is shown in Fig.~\ref{fig:device}b; the structure is similar to that used in previous AD experiments~\cite{DiLuca2026Aug_Nat}. The top graphite gate is shown in purple, the exposed hBN regions, where the top graphite has been etched away, are shown in blue, and the metallic gates are shown in orange.

We tune the AD following the procedure described in Ref.~\cite{DiLuca2026Aug_Nat}. The BG defines the AD potential, while the top gate (TG) sets the bulk filling factor to an integer quantum Hall (IQH) state and drives the oscillations. The side gates are set so that the system starts in the weakly coupled regime and are kept fixed for the rest of the experiment.\\

\begin{figure*}[tb!]
\centering
 \includegraphics[width = 0.95\textwidth]{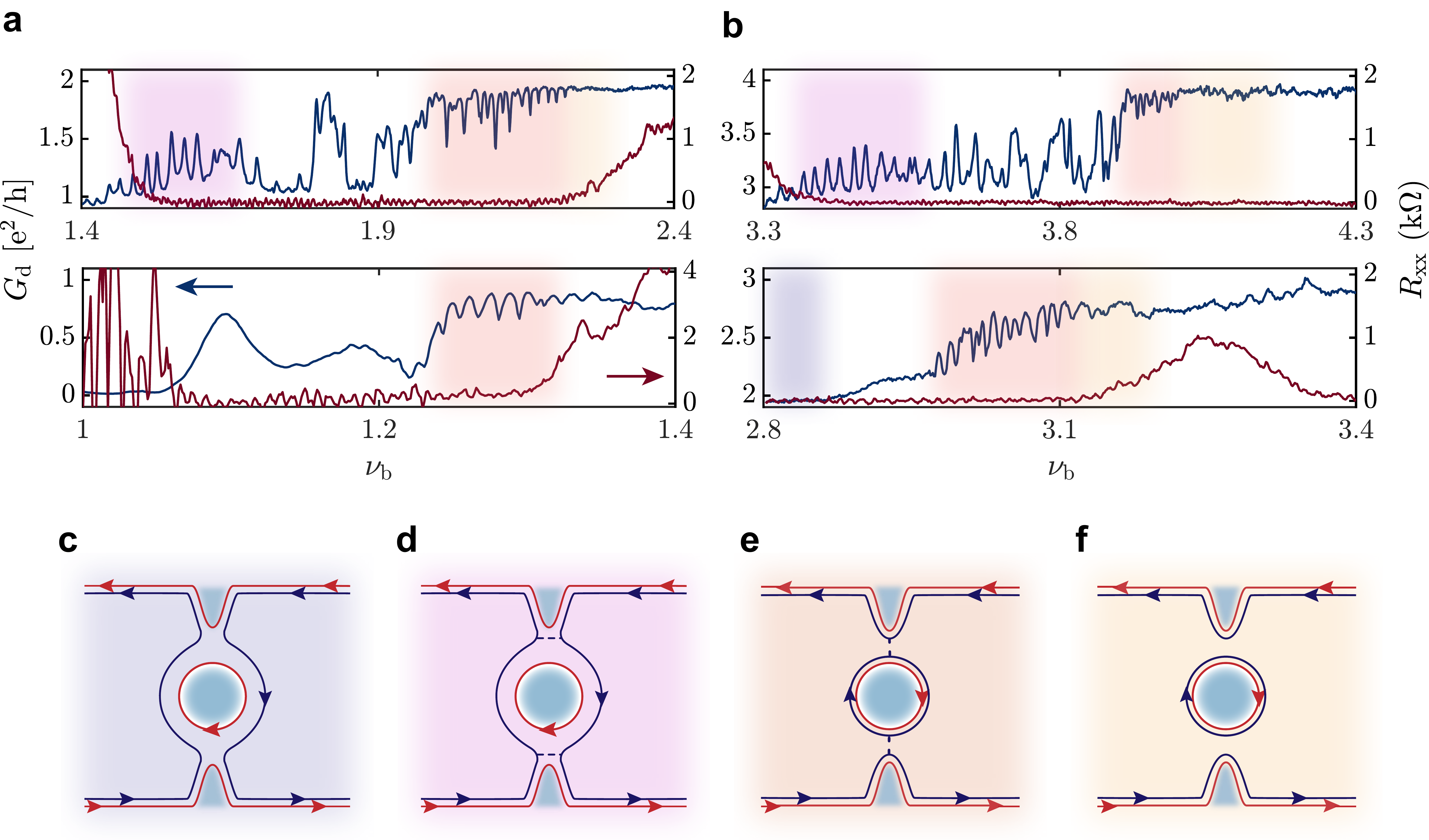}
 \caption{\textbf{AD dependence on $\nu_\mathrm{b}$.} Diagonal conductance $G_\mathrm{d}$ (blue) and longitudinal resistance measured outside the antidot, $R_\mathrm{xx}$ (red), as a function of the bulk filling factor $\nu_\mathrm{b}$, controlled by the TG, at $B = \qty{5}{\tesla}$. The back-gate voltage is set to $V_\mathrm{bg} = \qty{-1}{\volt}$ for $\nu = 1$, $\qty{-0.6}{\volt}$ for $\nu = 2$, $\qty{-0.1}{\volt}$ for $\nu = 3$ and $\qty{0}{\volt}$ for $\nu = 4$. \textbf{a}, For $\nu = 1$ (bottom panel) and $\nu = 2$ (top panel), and \textbf{b}, for $\nu = 3$ (bottom panel) and $\nu = 4$ (top panel). The measurements were taken at a fixed BG voltage by sweeping only the TG. The blue and red arrows in the bottom panel of (a) indicate the axis of the corresponding line scan. The bottom schematics \textbf{c-f} show the configuration of the edges in the respective regions. For simplicity, only the two inner edges are shown. We have used the same color code for $\nu = 1$, where in reality there is only one edge. \textbf{f}, Schematic of the situation in which the AD is not coupled to the extended edge states, as the inner edge is fully transmitted. \textbf{e}, The AD is weakly coupled to the inner extended edge state. \textbf{d}, The AD is strongly coupled to the extended edge state, as the inner edge is almost fully reflected. \textbf{c}, The AD is decoupled from the extended edge states as the inner edge is fully reflected while the outer is fully transmitted.} 
 \label{fig:nu_b}
\end{figure*}

We begin by studying how the constriction filling factor $\nu_\mathrm{c}$ controls the coupling regime of the system, in particular the transition from weak to strong coupling between the AD and the extended edge states, accessed by tuning the TG alone. The quantity that sets the coupling to the extended edges is $\nu_\mathrm{c}$, but we do not have direct experimental access to it, apart from the plateaux in the diagonal conductance. We therefore report the measurements as a function of the bulk filling factor $\nu_\mathrm{b}$, keeping in mind that $\nu_\mathrm{c} \leq \nu_\mathrm{b}$ and $\nu_\mathrm{b} \propto V_\mathrm{tg}$. In the following, $\nu_\mathrm{b}$ denotes the bulk filling factor, a continuous quantity proportional to the density, while $\nu$ denotes the quantized filling of the LL plateau containing $\nu_\mathrm{b}$. Figure~\ref{fig:nu_b}a--b shows the diagonal conductance $G_\mathrm{d}$ (blue) and the longitudinal resistance outside the AD $R_\mathrm{xx}$ (red) as a function of $\nu_\mathrm{b}$, for $\nu = 1$ and $\nu = 2$ in panel \textbf{a}, and for $\nu = 3$ and $\nu = 4$ in panel \textbf{b}. The measurements were taken at fixed BG voltage and at $B = \qty{5}{\tesla}$. We identify four distinct regions, highlighted in different colors, corresponding to the four configurations sketched in Fig.~\ref{fig:nu_b}c--f.

At each filling factor, $G_\mathrm{d}$ starts from a plateau close to the bulk value, $\nu_\mathrm{c} \simeq \nu_\mathrm{b}$. In this region, the AD is decoupled from the extended edge states, as the innermost edge is fully transmitted past the AD. This region is highlighted in light yellow, with the corresponding schematic in Fig.~\ref{fig:nu_b}\textbf{f}. As $\nu_\mathrm{b}$ decreases, the constriction filling factor decreases as well, with $\nu_\mathrm{c} < \nu_\mathrm{b}$, while the bulk remains in the integer quantum Hall regime ($R_\mathrm{xx} = 0\, \Omega$). $G_\mathrm{d}$ starts dropping from the plateau, and a current path opens through the AD, connecting the biased and grounded edges. This is the weakly coupled regime, where CD oscillations can be observed~\cite{Kou2012Jun, DiLuca2026Aug_Nat}. This region is highlighted in light orange, with the corresponding schematic in Fig.~\ref{fig:nu_b}e.

As $\nu_\mathrm{b}$ is further decreased, so is $\nu_\mathrm{c}$, and the innermost edge is increasingly reflected while the AD becomes more strongly coupled to the extended edges; we refer to this as the strongly coupled regime. This region is highlighted in light purple, with the corresponding schematic in Fig.~\ref{fig:nu_b}d. Eventually, the innermost edge is fully reflected while the outer edge is fully transmitted, and the system reaches $\nu_\mathrm{c} = \nu - 1$. The AD is again decoupled from the extended edge states. This configuration is visible in Fig.~\ref{fig:nu_b}b, bottom panel, where the bulk is still at $\nu_\mathrm{b} = 3$ ($R_\mathrm{xx} = 0\, \Omega$) while the diagonal conductance plateaus at $\nu = 2$. This region is highlighted in light blue, with the corresponding schematic in Fig.~\ref{fig:nu_b}c.

We now focus on the oscillations that appear in the strongly coupled regime (light purple), when the inner edge is almost fully reflected, and try to understand their origin. The first observation is that these oscillations appear only at even filling factors ($\nu = 2, 4$), whereas at odd filling factors ($\nu = 1, 3$) only the weakly-coupled oscillations are observed. For AB interference, the relevant energy is the thermal damping scale $T_0 = \hbar v / (k_B L)$, where $v$ is the edge velocity and $L$ the interferometer perimeter~\cite{Nakamura2019Jun}. The visibility decays as $e^{-T/T_0}$, so AB oscillations are resolvable only when $T_0$ is sufficiently large. In BLG the edge velocity depends on the filling factor and can differ between orbital sub-LLs~\cite{Fu2023Jan}, which provides a plausible mechanism for the even-$\nu$ selectivity, but we do not have an independent measurement of the drift velocity at each filling in our device to confirm this effect.

\begin{figure}[tb!]
\centering
 \includegraphics[width = 0.46\textwidth]{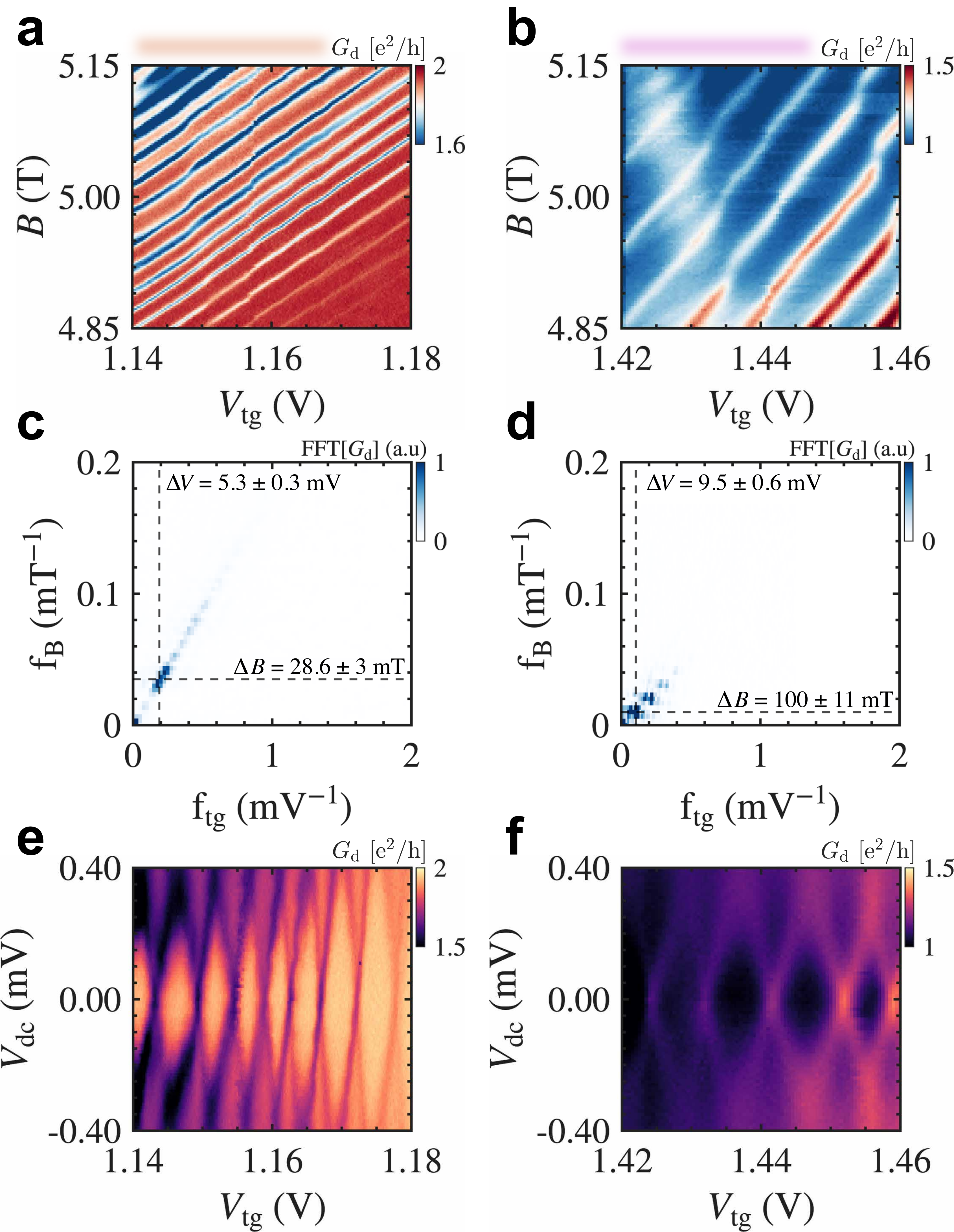}
\caption[Weakly and strongly coupled regimes at $\nu = 2$.]{\textbf{Weakly and strongly coupled regimes at $\nu = 2$.} \textbf{a-b}, Diagonal conductance $G_\mathrm{d}$ as a function of the magnetic field and the TG voltage at $\nu = 2$, in the \textbf{a}, weakly coupled and \textbf{b}, strongly coupled regimes. \textbf{c-d}, 2D-FFT of the corresponding oscillations. \textbf{e-f}, Diagonal conductance $G_\mathrm{d}$ as a function of the TG voltage and the DC bias.}
 \label{fig:nu2}
\end{figure}

\begin{figure}[tb!]
\centering
 \includegraphics[width = 0.46\textwidth]{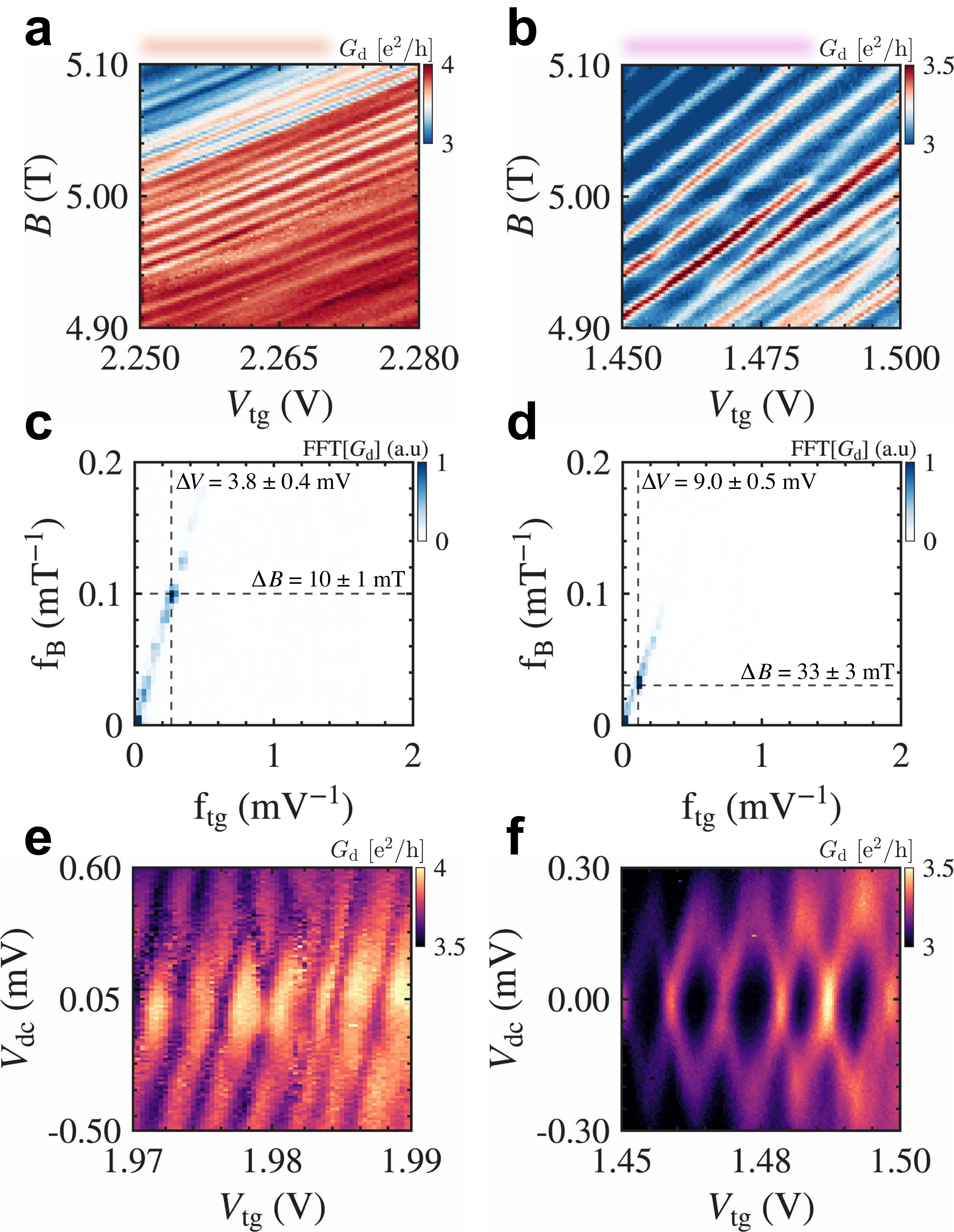}
\caption[Weakly and strongly coupled regimes at $\nu = 4$.]{\textbf{Weakly and strongly coupled regimes at $\nu = 4$.} \textbf{a-b}, Diagonal conductance $G_\mathrm{d}$ as a function of the magnetic field and the TG voltage at $\nu = 4$, in the \textbf{a}, weakly coupled and \textbf{b}, strongly coupled regimes. \textbf{c-d}, 2D-FFT of the corresponding oscillations. \textbf{e-f}, Diagonal conductance $G_\mathrm{d}$ as a function of the TG voltage and the DC bias.}
 \label{fig:nu4}
\end{figure}

Figure~\ref{fig:nu2} and Fig.~\ref{fig:nu4} report measurements in the two regimes at $\nu = 2$ and $\nu = 4$. The oscillations in panel (a) of both figures are those which we refer to as the weakly coupled regime: they appear as dips in the diagonal conductance and correspond to the CD regime. The oscillations highlighted in light purple, corresponding to the light purple region in Fig.~\ref{fig:nu_b}, are clearly different in character. They appear as peaks instead of dips, so a path through the AD now increases the conductance. We refer to this as the strongly coupled regime. From the 2D-FFT (Fig.~\ref{fig:nu2}d and Fig.~\ref{fig:nu4}d) we see that in both cases the magnetic field period and the gate voltage period are larger than in the weakly coupled regime. Phase slips are also visible in the strongly coupled oscillations. The DC-bias dependence is reported in Fig.~\ref{fig:nu2}f and Fig.~\ref{fig:nu4}f.\\

\begin{figure}[tb!]
\centering
 \includegraphics[width = 0.48\textwidth]{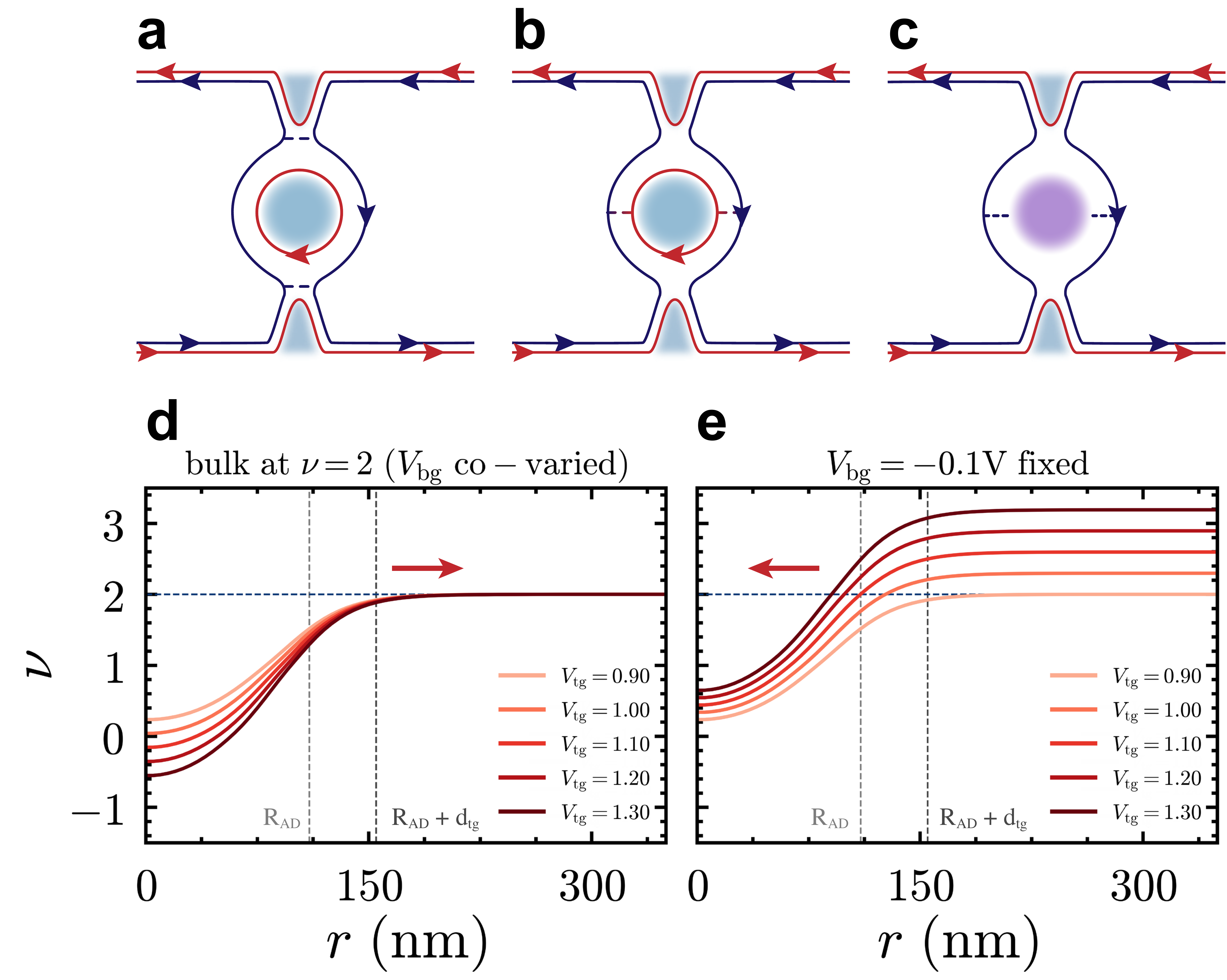}
\caption[Possible scenarios for the strongly coupled regime.]{\textbf{Possible scenarios for the strongly coupled regime.} In all cases, the inner edge (blue) is almost fully reflected while the outer edge (red) is almost fully transmitted. \textbf{a}, The AD is in the AB regime, with the system behaving as an interferometer in which the inner edge is partitioned twice before and after the AD. \textbf{b}, The inner edge interacts with the outer edge around the AD. \textbf{c}, The inner edge interacts with a puddle of charge in place of the AD. \textbf{d--e}, Simulated AD potential profile as a function of the distance from the center for different TG voltages, with \textbf{d}, the bulk filling factor held constant at $\nu = 2$, and \textbf{e}, the BG voltage held constant at $V_\mathrm{bg} = \qty{-0.1}{V}$. The arrows indicate the direction in which the AD area changes as $V_\mathrm{tg}$ increases. The two vertical lines mark the radius of the hole in the TG, $R_\mathrm{AD}$, and $R_\mathrm{AD} + d_\mathrm{t}$, where $d_\mathrm{t}$ is the top hBN thickness.}
 \label{fig:cases_3}
\end{figure}

Spin-selective measurements have shown that in AD only the inner edge contributes to the oscillations, while the outer edge does not participate~\cite{Kataoka2003Oct}. Hence, these oscillations have to come from the inner edge too. Figure~\ref{fig:cases_3} illustrates three possible scenarios for the strongly coupled regime where the inner edge is partially reflected while the outer edge is fully transmitted. 

Case (a) corresponds to AB oscillations of the inner edge. Geometrically, in this configuration the inner edge is partitioned twice before and after the AD and the system effectively becomes a FPI.

Case (b) describes a situation in which the inner edge interacts with the outer edge around the AD. The nature of this interaction can be of two types. The first is tunneling between the two edges, which, however, would require electrons to undergo a spin, valley or orbital flip, and is therefore energetically expensive. The second is a purely capacitive coupling between the two innermost co-propagating edges and can give rise to oscillations of the type observed here. Similar features have been reported in STM measurements around graphene antidots~\cite{Walkup2020Jan} and in strongly coupled edge channels in a graphene FPI~\cite{Werkmeister2024Aug, Yang2023Dec} and can be captured by a double-dot description. We note, however, that this picture sits at odds with the AD theory, where strong coupling between edge states encircling the AD defines the CD regime rather than a departure from it~\cite{LevySchreier2016Aug}. We therefore consider case (b) unlikely. 

Finally, case (c) represents the case in which the AD potential is substituted by a puddle of charge, similar to the electrostatic reconstruction observed in quantum dots in the QH regime~\cite{Roosli2021May, Roosli2020Mar}, and to some schematics used in some AD papers~\cite{Moreau2022Mar}. In our device, although we cannot directly access the AD filling factor, the AD density is mainly controlled by the BG and $\nu_\mathrm{AD}$ should be either on the hole side or close to $\nu = 0$. In either case, it differs from $\nu_\mathrm{b}$, so edge states must exist around the AD~\cite{Walkup2020Jan}, and we discard case (c) as well. So we believe that case (a) is therefore the most plausible scenario. We argue that the strongly coupled oscillations are AB oscillations of the inner edge. 

The slope of the equiphase lines provides a consistency check. In the AB regime, oscillations should be caused only by changes in area or magnetic field, not by changes in density. Finite elements numerical simulations reported in Fig.~\ref{fig:cases_3}e, show that the AD area decreases with increasing $V_\mathrm{tg}$, because the bulk density rises while the AD density (set by $V_\mathrm{bg}$ alone) stays fixed, so the field lines extend further into the AD region and the effective AD shrinks. In the oscillations in Fig.~\ref{fig:nu2}b, we observe that the equiphase lines follow a positive slope in the $(V_\mathrm{tg}, B)$ plane, meaning that an increase in $V_\mathrm{tg}$ corresponds to an increase in $B$ at constant phase, i.e.\ a decrease in area. This is consistent with the AB.

In the AB regime, the magnetic field period is $\Delta B_\mathrm{AB} = \phi_0/A$, while in the CD regime at filling $\nu$ the same area gives $\Delta B_\mathrm{CD} = \phi_0/(\nu A)$, so that the ratio $\Delta B_\mathrm{AB}/\Delta B_\mathrm{CD}$ is expected to equal $\nu$ for pure AB at constant area. Additionally, the corresponding gate-voltage prediction is $\Delta V_\mathrm{AB}/\Delta V_\mathrm{CD} = \nu$~\cite{LevySchreier2016Aug}. 

For the oscillations in the two regimes, we use the subscripts $s$ and $w$ to refer to the strongly and weakly coupled regimes, respectively.

At $\nu = 2$ we measure a gate period $\Delta V_\mathrm{s} = \qty{9.5 \pm 0.6}{\mV}$, giving $\Delta V_\mathrm{s}/\Delta V_\mathrm{w} = 1.81 \pm 0.08$, close to the expected value of $2$. However, from the magnetic field period $\Delta B_\mathrm{s} = \qty{100 \pm 11}{\mT}$ assuming pure AB ($\Delta B_\mathrm{s} = \phi_0/A$) we extract a diameter $D_\mathrm{AD} = \qty{230 \pm 14}{\nm}$. This diameter is smaller than the value $D \approx \qty{300}{\nm}$ obtained in the CD regime, but still larger than the lithographic AD ($\approx \qty{200}{\nm}$), so it is not unphysical. This indicates that the enclosed area in the AB regime is smaller than in the CD regime. A change in the area could also account for the deviation of the gate-period ratio from the expected value.

Analogous measurements at $\nu = 4$ give $\Delta V_\mathrm{s} = \qty{9.0 \pm 0.5}{\mV}$, $\Delta V_\mathrm{s}/\Delta V_\mathrm{w} = 2.37 \pm 0.12$, and $\Delta B_\mathrm{s} = \qty{33 \pm 3}{\mT}$, yielding a flux periodicity of $\phi_\mathrm{s}/\phi_0 = 1.21 \pm 0.14$. The departure from $\nu = 4$ in the gate ratio is larger than at $\nu = 2$, suggesting that the system at $\nu = 4$ sits in an intermediate regime between CD and AB rather than fully in the AB limit. The presence of phase slips in Fig.~\ref{fig:nu4}, where the diagonal-oscillation intensity locally drops, is consistent with this picture: similar phase slips at the boundary between different inter-edge coupling regimes have been reported in graphene FPIs~\cite{Werkmeister2024Aug, Yang2023Dec}.

Finally, the DC bias dependence (Fig.~\ref{fig:nu2}f and Fig.~\ref{fig:nu4}f) reveals features that look like Coulomb diamonds with lower conductance inside than outside, in contrast to the weakly coupled regime. However, these are not a smoking gun against the AB interpretation. AB oscillations in an antidot can display CD-like patterns in DC bias under certain conditions, as shown in Ref.~\cite{Moreau2022Mar}. The DC bias dependence alone cannot therefore discriminate between the two regimes.\\

In conclusion, we have shown that varying the constriction filling factor drives the quantum Hall antidot from a weakly coupled to a strongly coupled regime. At $\nu = 2$ the magnetic field periodicity changes from $\phi_0/\nu$ to $\phi_0$ and the gate voltage periodicity doubles, evidence that the oscillations in the strongly coupled regime are AB oscillations rather than CD; at $\nu = 4$ the system instead sits in an intermediate regime between the two. These results establish the AD as a geometry that can be electrostatically tuned across both the CD and AB dominated regimes, avoiding the bulk-to-edge coupling that limits conventional FPIs. This tunability makes the bilayer graphene antidot a platform for future studies of quasiparticle interactions and anyonic interference in the fractional quantum Hall regime.

\bibliography{biblio}

\section*{METHODS}

\subsection{Sample fabrication}

The device was fabricated using a standard van der Waals dry-transfer technique. First, graphene and hBN were mechanically exfoliated from bulk crystals on a SiO2/Si substrate. The desired flakes are identified under optical microscope and atomic force microscopy to check for any flake impurity. The stack was assembled using homemade poly(bisphenol A carbonate)/polydimethylsiloxane (PC/PDMS) to pick up all the flakes at a temperature $\sim$ 90 $^\circ$C. The stack was then transferred onto a doped silicon substrate with a 285 nm thick layer of thermally grown SiO$_2$ by melting the PC at a temperature of $\sim$ 180 $^\circ$C. The stack reported here has a 56 nm thick top hBN and a 33 nm thick bottom hBN layer. 

Device patterning was achieved through multiple steps of electron beam lithography, followed by reactive ion etching and metal deposition. First, the TG contacts were created by depositing an 18 nm Pd layer. The device geometry was then defined through reactive ion etching, initially using O$_2$, followed by SF$_6$, and then another O$_2$ step. The electrode pattern was subsequently created by etching with SF$_6$ and O$_2$, followed by angled deposition of a 5/100 nm Cr/Au layer with rotation.

Finally, the TG was refined with 30-second O$_2$ etch steps. During each step, the two-probe resistance between each gate bridge-contact was monitored to ensure that all gates were fully separated while minimizing etching of the hBN and achieving narrow line widths \cite{Ronen2021May}.\\

\subsection{Measurements}
The device was measured in a BlueFors dry dilution refrigerator with a base temperature of $\sim$ 10 mK. Electronic filters are installed on all transport lines to help thermalize electrons. Unless otherwise noted, a constant 5 T perpendicular magnetic field was applied. Resistance measurements are conducted using ZI lock-in amplifiers with a 1 V AC voltage excitation at a frequency of 33.333 Hz applied to a homemade voltage divider. A Yokogawa GS200 sets the DC bias. The current is measured through a current-to-voltage amplifier (K-tip variable gain transimpedance amplifier), and voltages are amplified at room temperature by 100 before being sent to the lock-in amplifier. For DC bias measurements, a homemade voltage adder is employed. Yokogawa GS200 is used to apply voltage to each gate. To improve the quality of the ohmic contacts, a positive voltage of $V_\mathrm{Si} = 14.5$ V is applied to the silicon gate, doping the graphene layer next to the contacts in a highly electron-doped region. All measurements are performed in the electron-doped regime. 

The devices have ohmic contacts along the path of both upstream and downstream extended edges, allowing for current injection and differential chemical potential measurements across the AD. Unless otherwise stated, all measurements were performed with an applied voltage of V$_\mathrm{ac} = 10$ $\mu$V for the IQH regime, while measuring the transmitted current, $I_\mathrm{t}$. Furthermore, longitudinal resistance $R_\mathrm{xx} = V_\mathrm{xx}/I_\mathrm{t}$ is measured outside the AD and the transmitted resistance across it $R_\mathrm{t} = V_\mathrm{t}/I_\mathrm{t}$, while the measurement of the diagonal voltage drop gives the diagonal conductance through the AD, $G_\mathrm{d} = I_\mathrm{t}/(V_\mathrm{d}) \equiv \nu_\mathrm{c} e^2/h$, where $\nu_\mathrm{c}$ is the constriction filling factor.

\textbf{\begin{center}Author contributions\end{center}}
M.B. and M.D.L. conceived the project. M.B. supervised the project. M.D.L. fabricated the devices with inputs from Z.Z and T.F. M.D.L. performed the measurements and analyzed the data with input from E.H. K.W. and T.T. provided the hBN crystals. M.D.L., E.H., F.K. and M.B. wrote the manuscript with input from all authors.

\begin{acknowledgments}
The authors acknowledge M. Heiblum, T. Ihn and S. H. Simon for valuable discussions. M.D.L thanks M. Coraiola for the support in the COMSOL simulations. M.D.L, E.H and Z.Z acknowledge funding from SNSF. M.B. acknowledges the support of SNSF Eccellenza grant No. PCEGP2\_194528, and support from the QuantERA II Programme that has received funding from the European Union’s Horizon 2020 research and innovation program under Grant Agreement No 101017733. 
F.K. acknowledges support from the European Research Council (ERC) as part of the project NONLOCAL under grant agreement No 856526. K.W. and T.T. acknowledge support from the JSPS KAKENHI (Grant Numbers 20H00354 and 23H02052) and World Premier International Research Center Initiative (WPI), MEXT, Japan. M.D.L acknowledges Blender for its open source software.
\end{acknowledgments}

\textbf{\begin{center}Data availability\end{center}}
The data supporting the findings of this study are available from the corresponding author upon reasonable request.

\clearpage
\onecolumngrid

\section*{Supplementary Material}

\subsection*{Device Characterization}

For the device characterization refer to AD\#1 in the Supp. Information of Ref~\cite{DiLuca2026Aug_Nat}.

\subsection*{Oscillations in the strongly coupled regime at $\nu = 2$}

\begin{figure*}[htb!]
\renewcommand{\thefigure}{S1}
 \includegraphics[width = 0.95\textwidth]{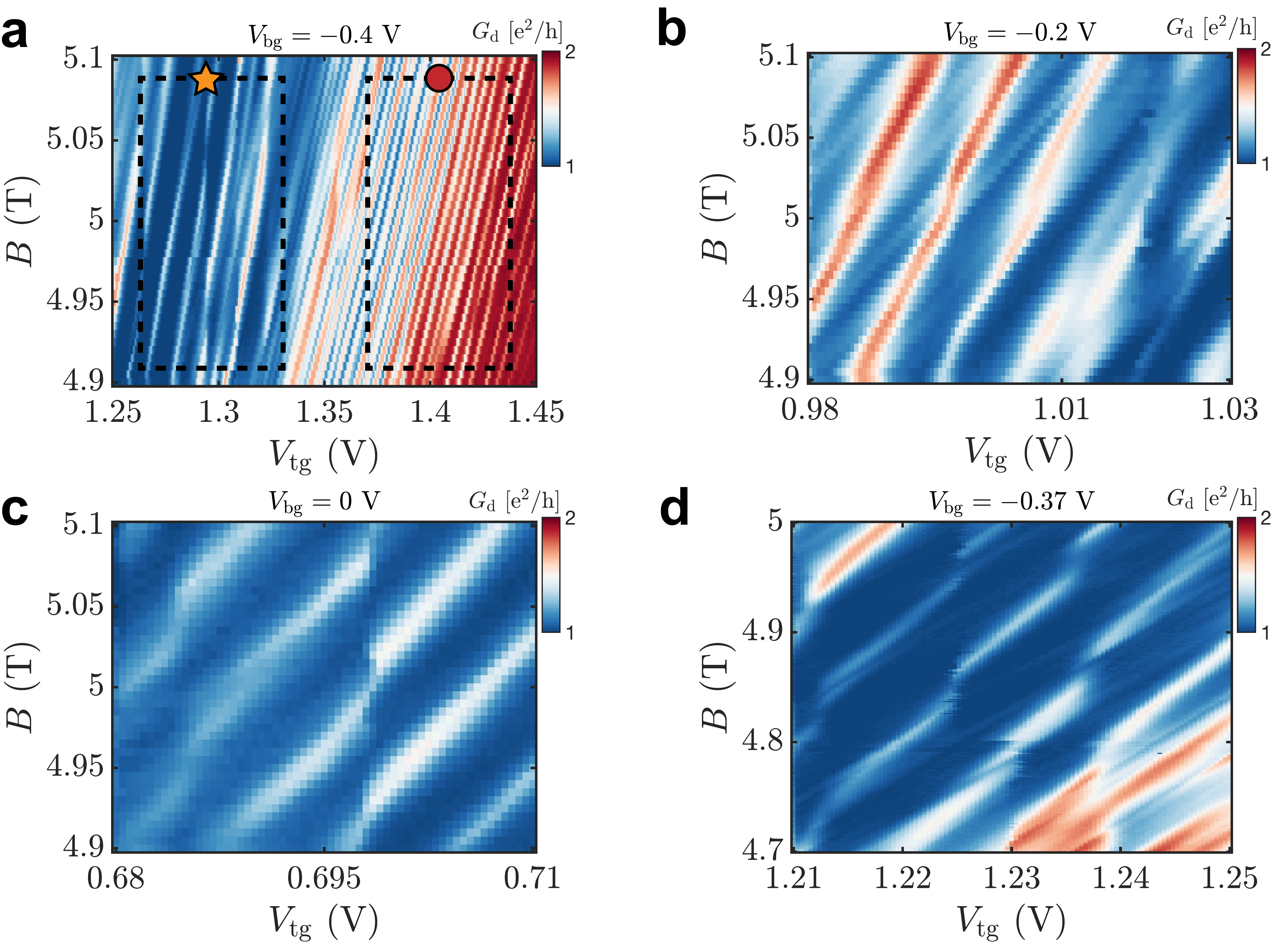}
 \begin{center}
 \caption{\textbf{Supplementary oscillations at $\nu = 2$ in the strongly coupled regime} \textbf{a}, Diagonal conductance oscillations at $B = 5$\,T in function of the top gate and the magnetic field over a large gate range. The oscillations continuously transition from the weakly coupled regime marked by the red dot to the strongly coupled, marked by the yellow star. \textbf{b}, Diagonal conductance oscillations in a gate voltage where both kinds of oscillation are visible.\textbf{c-d}, Diagonal conductance oscillations in the strongly coupled regime for two different back gate values.} 
 \label{fig:sup_5_b}
 \end{center}
\end{figure*}

We report here supplementary oscillations of the strongly coupled regime for $\nu = 2$ at B = 5 T. Figure \ref{fig:sup_5_b}a shows a large scan that goes from the weakly coupled regime marked by the red dot characterized by $\phi_0/2$ and $e$ oscillations, to the strongly coupled, marked by the yellow star, characterized by $3\phi_0/2$ and $2e$ oscillations. The transition between the two regimes is continuous and and in the middle region shown in figure \ref{fig:sup_5_b}b (for a different back gate voltage), it is possible to observe the superposition of the two oscillations modes. Finally \ref{fig:sup_5_b}c--d show the oscillations in the strongly coupled regime for different back gate values. 

\newpage

\subsection*{Oscillations in the strongly coupled regime at $\nu = 4$}

\begin{figure*}[htb!]
\renewcommand{\thefigure}{S2}
 \includegraphics[width = 0.95\textwidth]{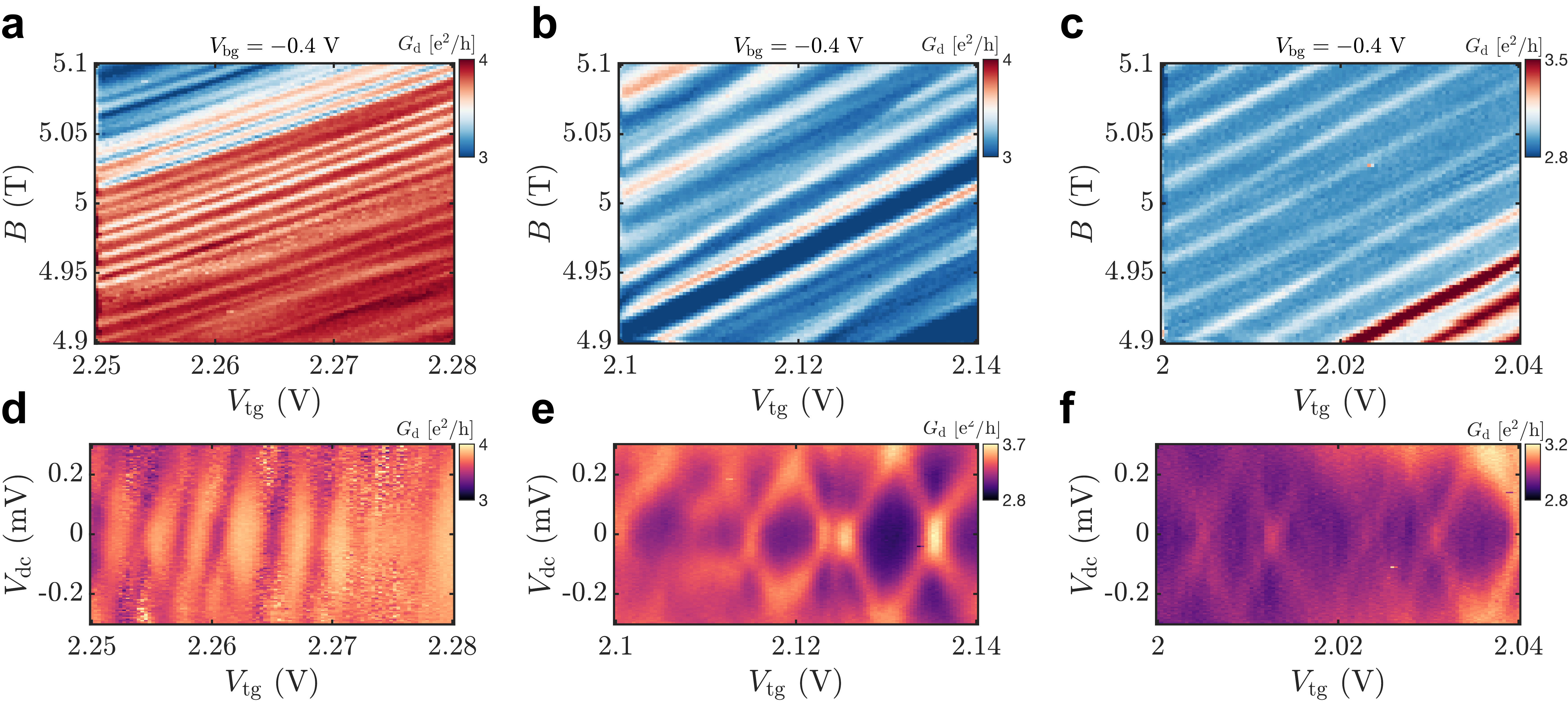}
 \begin{center}
 \caption{\textbf{Supplementary oscillations at $\nu = 4$ } \textbf{a-c}, Diagonal conductance oscillations at $\nu = 4$ at $B = 5$\,T in function of the top gate and the magnetic field at different back gate voltages.} 
 \label{fig:sup_5_a}
 \end{center}
\end{figure*}

We report here supplementary oscillations from the weakly coupled to the strongly coupled regime for $\nu = 4$ at $B = 5$\,T. The oscillations in figure \ref{fig:sup_5_a}b--c exhibit a larger period, as explained in the main text. 

\subsection*{Gate Voltages used for figures in the main text}

In Table~\ref{tab:volatges} we report the gate voltages corresponding to figures in the main text. When the voltage was the sweeping one it is not indicated. 

\begin{table}[htb!]
 \centering
 \begin{tabular}{c|c|c}
 Figure & V$_\mathrm{bg}$ (V) & V$_\mathrm{tg}$ (V)\\ \hline

Fig. 2a $\nu = 1$ & 1 & / \\
Fig. 2a $\nu = 2$ & -0.6 & / \\
Fig. 2b $\nu = 3$ & -0.1 & / \\
Fig. 2b $\nu = 4$ & 0 & / \\

Fig. 3a& 0 & / \\
Fig. 3b& -0.6 & / \\
 
Fig. 4a& -0.4 & / \\
Fig. 4b & 0 & / \\
 
 \end{tabular}
 \caption{\textbf{Fixed voltages for main figures}}
 \label{tab:volatges}
\end{table}

\end{document}